\documentclass[preprint,aps,amsmath,amssymb]{revtex4}
\usepackage{graphicx}
\usepackage{dcolumn}
\usepackage{bm}
\usepackage[bookmarks,hypertex]{hyperref}

\begin{document}

\title{\textit{Ab Initio} Study on Electronic Structure of ZrB$_{12}$ under High Hydrostatic Pressure}

\author{Z. F. Hou}
\affiliation{Department of Physics,
Fudan University, Shanghai, P. R. China 200433}
\email{zfhou@fudan.edu.cn} 
\date{\today}

\begin{abstract}
Using projector augmented wave approach within the generalized gradient approximation, we have studied the structural property and electronic structure of ZrB$_{12}$. The calculated lattice constants and bulk modulus are in good agreement with the available experimental values. A detailed study of the electronic structure and the
charge-density redistribution reveals the features of strong covalent
B-B and weak covalent Zr-B bondings in ZrB$_{12}$.  The states at the Fermi level mainly come from the $\sigma$-$p_{y}$ and $\pi$-$p_{z}$ orbitals of boron atoms, which are slightly hybridized with the $t_{2g}$-$d_{xz}$ and $t_{2g}$-$d_{yz}$ orbitals of Zr atoms. As the increased hydrostatic pressure on ZrB$_{12}$,  the total density of states at the Fermi level decreases.

\end{abstract}
 \maketitle
\section{\label{sec:intro}Introduction}
Metal borides have been the subject of extensive research
in recent years~\cite{Buzea01,Canfield02,Ivanovskii03} because they exhibit unique properties such as high melting
points, high hardness values, and excellent oxidation
resistance. These particular properties mean borides are promising materials for
new heat-resistant, corrosion-resistant, and wear-resistant alloys
and coatings~\cite{Aizawa05}.  In particular, the discovery of superconductivity ($T_{C} \backsim$ 39 K) in magnesium diboride (MgB$_2$)~\cite{Nagamatsu01}, which is crystallized in a simple layered hexagonal AlB$_2$ structure, has stimulated a renewed  interest to search for new superconductors and to understand their mechanism of superconductivity in borides. 

ZrB$_{12}$ is found to have the highest superconductivity critial temperture ($T_C \sim$ 6 K) in the family of dodecaborides \textit{M}B$_{12}$~\cite{Matthias68,Tsindlekht04,Lortz05}. In order to understand its superconductivity, some efforts are made by measuring the specific heat, magnetic suseptibility, resistivity, thermal expansion, electron transport, penetration depth, and the upper critical magnetic field of ZrB$_{12}$~\cite{Gasparov04,Tsindlekht04,Daghero04,Khasanov05,Lortz05,Gasparov05,Wang05,Leviev05,Gasparov06,Tsindlekht06}. Specially, a negative pressure effect on $T_{C}$ of  ZrB$_{12}$ is observed in its magnetization measurements under pressure, which contrasts to the previous suggestion that the superconducting critical temperature of ZrB$_{12}$ might increase under pressure~\cite{Fisk69}.  The previous band structure calculations on ZrB$_{12}$ mainly concern the electronic structure of ZrB$_{12}$ under ambient pressure~\cite{Shein03,Jager06}.  Therefore, we perform \textit{ab initio} total energy calculations to study the properties of ZrB$_{12}$, epescially about its electronic structure under high pressure.   

\section{Method}
All calculations on ZrB$_{12}$ are performed
using the projector augmented wave (PAW) approach as implemented in the VASP
code \cite{GK1,GK2}. The exchange-correlation energy functional is
treated with the generalized gradient approximation (GGA)
with Perdew and Wang parameterization (known as
GGA-PW91)~\cite{JPJa}. Electron-ion interaction is represented by the PAW method \cite{Blochl94,Kresse99} and
wave functions are expanded by the plane waves up to an energy
cutoff of 500 eV. Brillouin-zone integrations are approximated
using the special {\em k}-point sampling of Monhkorst-Pack scheme
\cite{HJJD} with a 9$\times$9$\times$9 grid. The above calculation parameters of energy cutoff and $k$-mesh make the total energy
convergence to 2 meV/atom. The internal atomic coordinates of cubic ZrB$_{12}$  are
relaxed at a series of fixed volumes until the force exerting on each atom is less than 0.01 eV/\AA. The obtained energies are fitted with the third-order
Birch-Murnaghan equation of state (EOS) \cite{Birch47} to give the
equilibrium volume and the minimum energy. The final calculated cell
parameters are listed in Table~\ref{tab:1}.

\section{Results and Discussions}
The crystalline structure of ZrB$_{12}$ is cubic and belongs to space
group O$_{h}^5$-$Fm\overline{3}m$ (No.225) ~\cite{Kuzma83}, shown in Fig.~\ref{fig:1}. The Wyckoff positions of Zr and boron atoms are 4\textit{a} (0,
0, 0) and 48\textit{i} (1/2, \textit{x}, \textit{x}), respectively, i.e. the Zr atoms and cuboctahedral B$_{12}$ clusters are arranged as in the NaCl-like structure. The 13-atom fcc unit cell
of cubic ZrB$_{12}$ is characterized by single lattice constant and the internal parameter 
of boron atoms. In present work, we obtain the lattice constant \textit{a} is 7.409 \AA$ $ and the internal coordinate parameter \textit{x} of boron atoms is 0.161, in good agreement with experiment values~\cite{Kuzma83,Matkovic65} (See Table~\ref{tab:1}). Based on GGA calculations, the bulk modulus and its pressure
derivative for ZrB$_{12}$ are predicted as 233.0 GPa  and 3.64, respectively. Note that in the investigation of electron-phonon coupling depentent on pressure of ZrB$_{12}$ the authors~\cite{Khasanov05} used the bulk modulus of an analogy compound UB$_{12}$ ($B_{0}$ = 249 GPa) because the experimental value of bulk modulus of ZrB$_{12}$ is not reported. In addition, the predicted bulk modulus of ZrB$_{12}$ is close to that of ZrB$_2$ (245 GPa obtained from the ultrasonic measurements~\cite{Okamoto03} or 238 GPa predicted from \textit{ab initio} calculations~\cite{Milman05}) crystallized in hexagonal layered AlB$_2$-like structure.

The energy band structure and electronic density of states (DOS) of cubic
ZrB$_{12}$ in equilibrium are shown in Fig. \ref{fig:2}(a) and \ref{fig:3}, respectively. These may be
compared well with  previously calculated results of ZrB$_{12}$ based on a variety
of computational methods~\cite{Shein03,Jager06}. Overall these suggest that ZrB$_{12}$ exhibits a  metallic character with the mixed \textit{p}-\textit{d} states rather than a simple \textit{p}- or \textit{d}-metal~\cite{Khasanov05}. The valence bands in
Fig. \ref{fig:2}(a) are split into three
disjointed groups. The lowest group centered at -15 eV below the Fermi level ($E_{F}$) has a width of
about 1 eV and  the corresponding bands mostly come  from the bonding states of B-2$s$ orbitals. Combining from the DOS figures (\ref{fig:3}), the higher one in the energy range from -13 eV to -10 eV below the $E_{F}$ is largely contributed by the antibonding states of B-2$s$ orbitals.  The upper one in the energy range from
-10 eV to $E_{F}$ is mainly composed by B-2$p$ orbitals and Zr-4$d$ electrons. In addition, it  suggests that there is a weak hybridization between B-2$p$ and
Zr-4$d$ orbitals occuring below the $E_{F}$. The total density of states at the Fermi level $N(E_F)$ is about 1.478 states/eV$\cdot$unit cell, in consistence with the calculated value in other work~\cite{Shein03}. The states at $E_{F}$ is mainly made by the 2$p$ electrons of boron atoms ($\sigma$-$p_{y}$ and $\pi$-$p_{z}$ orbitals) and 4$d$ levels of Zr atom ($t_{2g}$-$d_{yz}$ and $t_{2g}$-$d_{xz}$ orbitals). This can also be seen from the the
partial charge density calculated within a 0.2 eV energy window
around $E_{F}$ for ZrB$_{12}$, which is shown in Fig.~\ref{fig:4}(b).

In order to understand the Zr-B and B-B bondings of ZrB$_{12}$ in equilibrium, the line charge densitiy along nearest
neighbor Zr$-$B and B$-$B pairs are illustrated in Fig.~\ref{fig:5}. The calculated bond lengths of atoms are $d_\text{Zr-B}$ = 2.75 \AA $ $ and $d_\text{B-B}$ = 1.68 \AA, which agree well with the experimental values~\cite{Matkovic65}. It is clearly seen that charge highly accumulates
around Zr atom due to its more valence electrons than that of boron atom.  Contrast to the strong covalent character of B-B bonding, the Zr-B bonding exhibits  a weak covalent character. It also supported by
above discussion that the $d$ orbital of Zr atoms and $p$ orbital of boron atoms have
a weak hybridization. In order to
further reveal the topology of the Zr-B and  B-B bondings, Fig.~\ref{fig:4}(b) illustrates the contour plots of electron density on crystallographic (010) plane.  It again indicates that B-B bonding exhibits a strong covalent character.   

Now we turn to discuss the effect of high hydrostatic pressure on electronic structure of ZrB$_{12}$. The values of DOS at the $E_{F}$ for ZrB$_{12}$ under a serial of hydrostatic pressure are listed in Table~\ref{tab:2}.  It can be seen that as the pressure increased from 0 to 22.1 GPa, the total density of states at the $E_{F}$ decreases about  0.264 states/eV$\cdot$unit cell. Considering the electron-phonon coupling constant $\lambda_\text{el-ph}$ depends linearly on the $N(E_F)$~\cite{McMillan68}, our calculated results indicate that the $\lambda_\text{el-ph}$ of ZrB$_{12}$ would decrease under the high pressure. It also supported by recent experimental observation~\cite{Khasanov05} that there is a negative pressure effect on $T_{C}$ and $\lambda_\text{el-ph}$ of ZrB$_{12}$ with $dT_{C}/dp$ of -0.0225 K/kbar and $d ln \lambda_\text{el-ph}/dp$ of -0.2\%/kbar. Additionaly, as the hydrostatic pressure increased, the contribution to $N(E_F)$ of Zr atoms decrease than that of boron atoms. This could be understood from that the B-B bonding is stronger than Zr-B bondingbond, and hence the bond length of B-B atoms decreases more slightly under the high pressure (see Table~\ref{tab:2}). Upon the hydrostatic pressure to 22.1 GPa, the width of valence bands of ZrB$_{12}$ is increased to -16 eV as shown in Fig.~\ref{fig:2}(b), companied with the 2$s$-bonding states of B boron atoms toward lower energy.

\section{Conclusions}

The structural property and electronic
structure of ZrB$_{12}$ have been calculated and analyzed. Our
calculated structural parameters including the lattice constants and
internal parameters are in good agreement with the experiment data.
The detailed study of
the electronic structure and charge density redistribution reveals
the strong covalent of bonding of B-B atoms and role of  \textit{p} orbitals of boron atoms and \textit{d} orbitals of Zr atom in the supercondutivity of ZrB$_{12}$.  Our calculated results also indicate that there is a negative pressure effect on the superconductivity of ZrB$_{12}$.

\section{Acknowledgements}
The author acknowledges support from Shanghai Postdoctoral Science
Foundation under Grant No. 05R214106. The computation was performed
at  Supercomputer Center of Fudan.

\clearpage
\newpage

\begin{table}
\caption{\label{tab:1}Calculated lattice parameter ($a$, in \AA), atomic coordinate parameter ($x$ for boron atoms), volumes of unit cell ($V_{0}$, in \AA$^3$), and bulk modulus ($B_{0}$, in GPa) of cubic ZrB$_{12}$, along a comparison with other theoretical work and available experiment values.}
\begin{ruledtabular}
\begin{tabular}{ccccc}
        & $a$  & $x$  &  $V_{0}$ & $B_{0}$ \\
	\hline
     This work  & 7.409  & 0.1696 &  101.69& 233.0 \\
     Other work~\cite{Shein03} & 7.419 &  - & 102.09&   -   \\
     Other work~\cite{Jager06} & 7.408 & 0.1696 & 102.09&   -   \\
     Expt.~\cite{Kuzma83} & 7.408 &  0.166 &    101.63& -   \\
     Expt.~\cite{Matkovic65} & 7.408 &  0.1699 &    101.63& -   \\
\end{tabular}
\end{ruledtabular}
\end{table}

\begin{table}
\caption{\label{tab:2} The bond lengths of atoms ($d_\text{Zr-B}$ and $d_\text{B-B}$, in \AA), total and partial density of states at the Fermi level ($N$($E_{F}$) and $N_{par}$($E_{F}$), in states/eV$\cdot$unit cell) of cubic ZrB$_{12}$ at different hydrostatic pressures.}
\begin{ruledtabular}
\begin{tabular}{cc cc cc }
Pressure &$d_\text{Zr-B}$ & $d_\text{B-B}$& $N$($E_{F}$) &  \multicolumn{2}{c}{$N_{par}$($E_{F}$)}\\
 &     &  &  & Zr-4\textit{d} & B-2\textit{p}   \\
 \hline
0 GPa  &  2.75& 1.68& 1.478  & 0.360 & 0.585\\
4.9 GPa  & 2.73 & 1.67& 1.306 & 0.315 & 0.518\\
10.2 GPa & 2.71& 1.66 & 1.251 & 0.298 & 0.503 \\
22.1 GPa & 2.68 & 1.64 & 1.214 &0.284 & 0.501 \\
\end{tabular}
\end{ruledtabular}
\end{table}

\begin{figure}
\includegraphics*[scale=0.4]{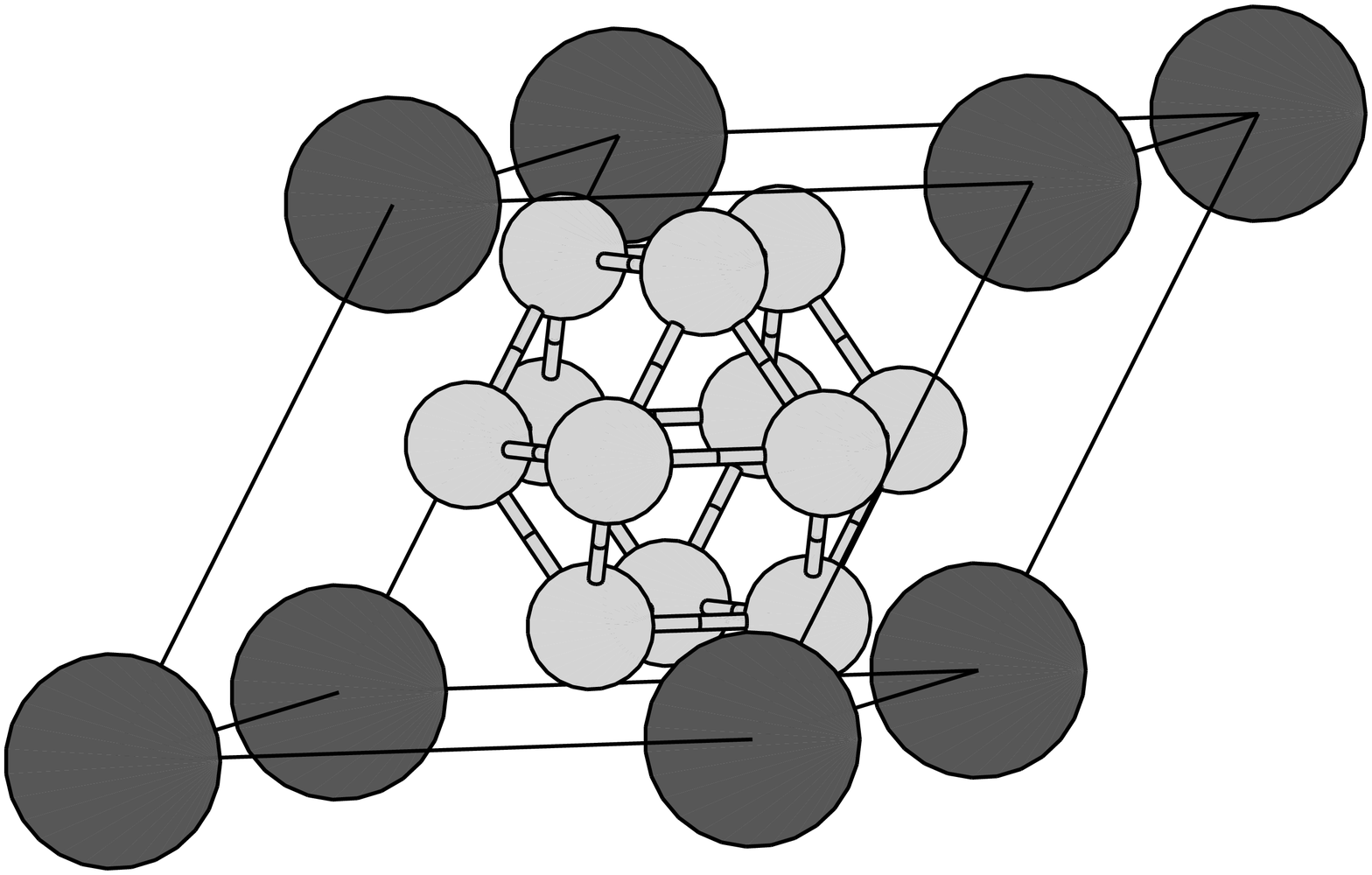}
\caption{\label{fig:1} }
\end{figure}

\begin{figure}
\begin{center}
\includegraphics*[scale=0.8]{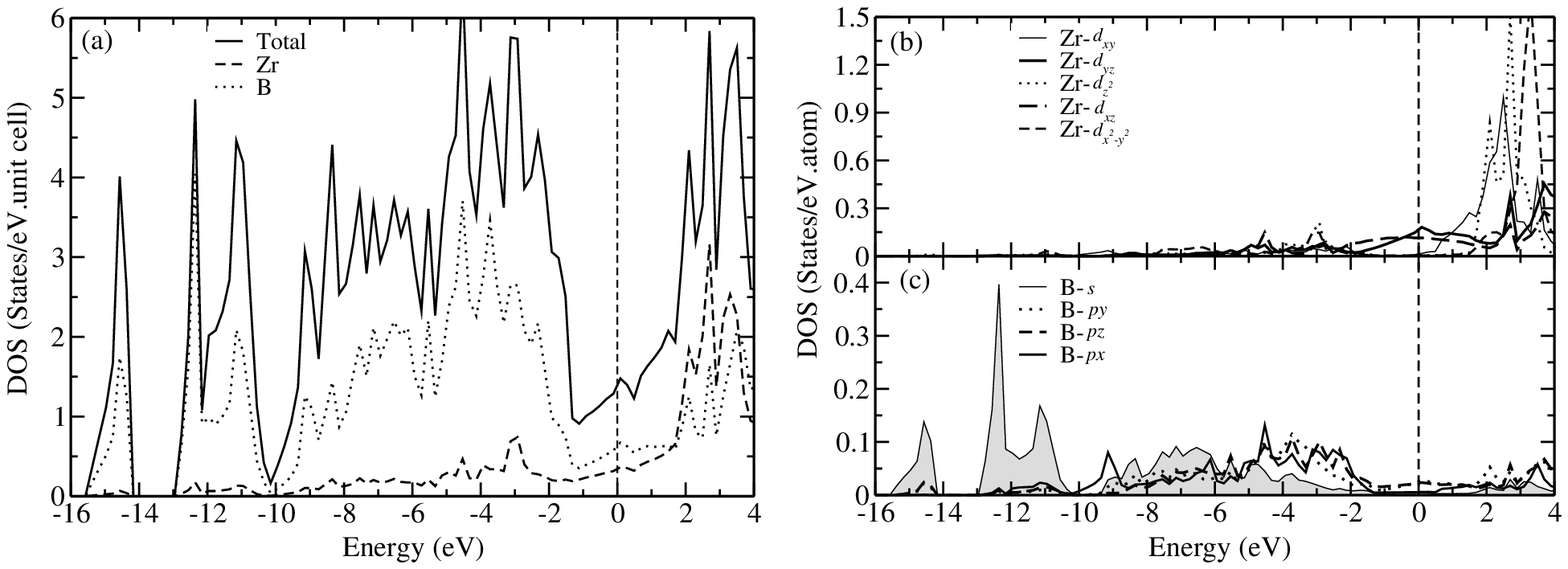}
\caption{\label{fig:3}}
\end{center}
\end{figure}

\begin{figure}
\begin{center}
\includegraphics*[scale=0.8,angle=0]{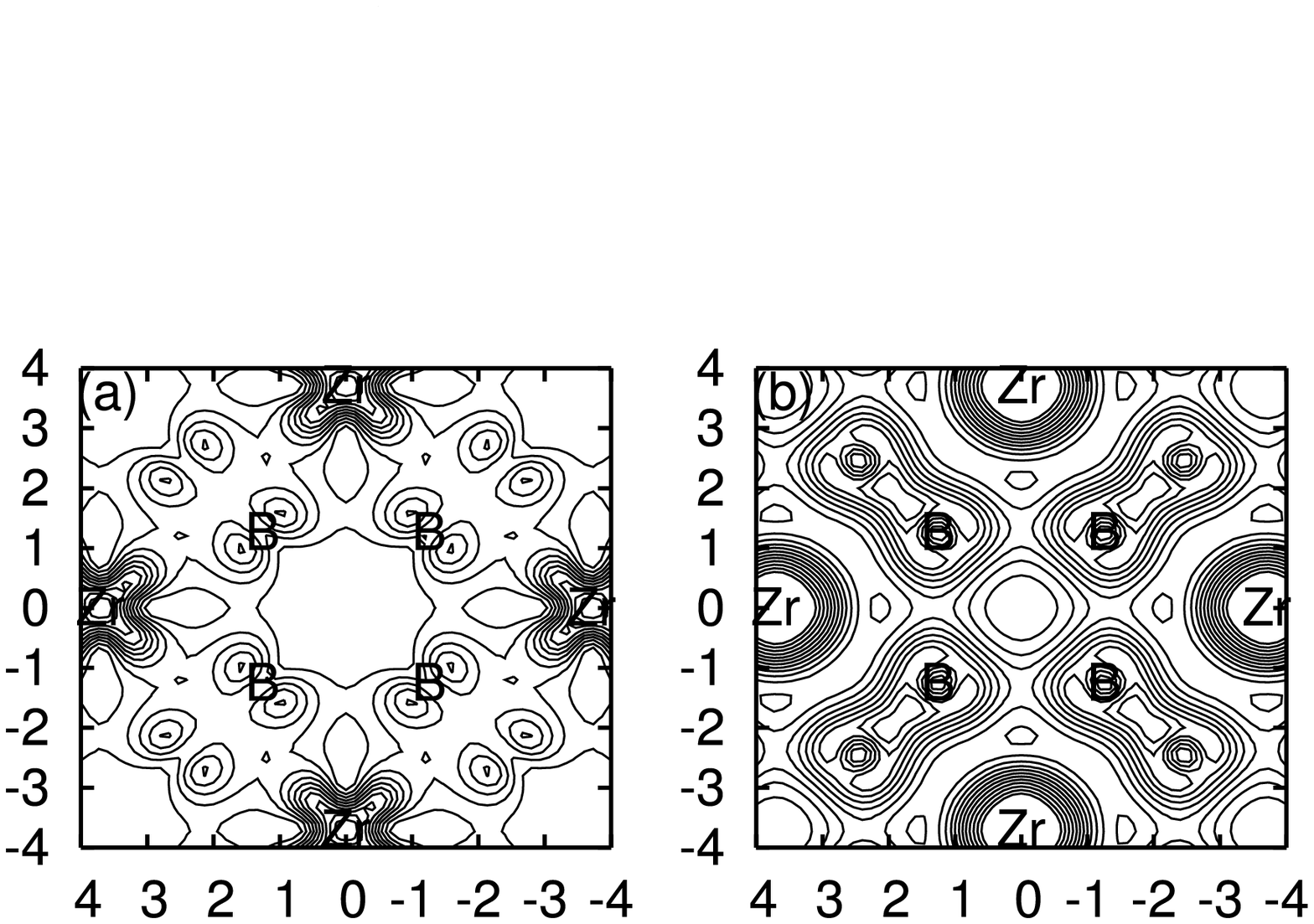}
\caption{\label{fig:4} }
\end{center}
\end{figure}

Figure Captions
\begin{itemize}
\item FIG \ref{fig:1}: The ball-and-stick model for unit cell of ZrB$_{12}$. Zr atoms are shown as dark gray balls and boron
atoms as light gray balls.

\item FIG \ref{fig:2}: Electronic band structure of ZrB$_{12}$ (a) in equilibrium and (b) at 22.1 GPa hydrostatic pressure. The zero of energy is set as the
Fermi level and shown in dash line.

\item FIG \ref{fig:3}: (a) Total and partial density of states (DOS) , (b) and (c) angular momentum decompsed DOS of ZrB$_{12}$ in equilibrium.

\item FIG \ref{fig:4}: Contour plots of the charge density on (010) surface of ZrB$_{12}$ in equilibrium: (a) partial charge density with the energy window of [$E_{F}-0.1$, $E_{F}+0.1$] is in an increment of 2.0$\times$10$^{-3}$ \textit{e}/\AA$^3$ from 0 to 2$\times$10$^{-2}$ \textit{e}/\AA$^3$ and (b) total charge density is in an increment of 0.1 \textit{e}/\AA$^3$ from 0 to 1.5 \textit{e}/\AA$^3$.

\item FIG \ref{fig:5}: Charge density at the equilibrium lattice parameter of ZrB$_{12}$, along the lines  strarting from: (a) Zr atom to nearst neighbor B atom, and (b) B atom to nearest neighbor B atom.

\end{itemize}


\begin{figure}
\includegraphics*[scale=0.6]{band.eps}
\caption{\label{fig:2} }
\end{figure}


\begin{figure}
\begin{center}
\includegraphics*[scale=0.6,angle=0]{charge-line.eps}
\caption{\label{fig:5} }
\end{center}
\end{figure}

\end{document}